\documentclass[12pt]{article}
\usepackage{amsfonts}
\usepackage{amssymb}
\usepackage{amsbsy}
\usepackage{graphics}

\input epsf.sty

\newlength{\TZ}
\newcommand{\BEQ}{\begin{equation}}     
\newcommand{\BEA}{\begin{eqnarray}}
\newcommand{\EEQ}{\end{equation}}       
\newcommand{\EEA}{\end{eqnarray}}
\newcommand{\vph}{\varphi}              
\newcommand{\D}{{\rm d}}                
\newcommand{\II}{{\rm i}}               
\newcommand{\wit}[1]{\widetilde{#1}}    
\newcommand{\lap}[1]{\overline{#1}}     

\renewcommand{\vec}[1]{\boldsymbol{#1}} 

\newcommand{\zeile}[1]{\vskip #1 \baselineskip} 

\newcommand{\appsection}[1]{\setcounter{equation}{0} \section*{Appendix. #1}
\renewcommand{\theequation}{A\arabic{equation}}
              \renewcommand{\thesection}{#1} }

\catcode`\@=11
\def\numberbysection{\@addtoreset{equation}{section}
        \def\theequation{\thesection.\arabic{equation}}}
\numberbysection

\begin{document}

\begin{titlepage}


\vskip 1.5 cm
\begin{center}
{\Large \bf Response of non-equilibrium systems with long-range 
initial correlations}
\end{center}

\vskip 2.0 cm
\centerline{  {\bf Alan Picone} and {\bf Malte Henkel} }
\vskip 0.5 cm
\centerline {Laboratoire de Physique des 
Mat\'eriaux,\footnote{Laboratoire associ\'e au CNRS UMR 7556} 
Universit\'e Henri Poincar\'e Nancy I,} 
\centerline{ B.P. 239, 
F -- 54506 Vand{\oe}uvre l\`es Nancy Cedex, France}

\begin{abstract}
\noindent 
The long-time dynamics of the $d$-dimensional spherical model with a 
non-conserved order parameter and quenched from an initial state with 
long-range correlations is studied through the exact calculation of the
two-time autocorrelation and autoresponse functions. In the aging regime,
these are given in terms of non-trivial universal scaling functions of both 
time variables. At criticality, five distinct types of aging are found, 
depending on the form of the initial correlations, while at low temperatures 
only a single type of aging exists. The autocorrelation and autoreponse 
exponents are shown to be generically different and to depend on the initial 
conditions. The scaling form of the two-time response functions agrees with
a recent prediction coming from conformal invariance. 
\end{abstract}
\end{titlepage}

\section{Introduction}

The slow dynamics of non-equilibrium systems displays several
characteristic features which are absent for systems in thermodynamic 
equilibrium, see \cite{Bray94,Bouc98,Cate00,Godr01} for reviews. 
Most notable among these are the break-down of the
fluctuation-dissipation theorem and the aging of the system as conveniently
displayed in the two-time autocorrelations and auto-response functions. 
While these features were first observed in glassy systems, they also
occur in simple ferromagnetic spin systems without disorder. The present
paper studies such relatively simple non-equilibrium systems. 

Usually, the system is prepared in some initial state (a completely 
disordered initial state of effective infinite temperature is common) and
then brought out of equilibrium even at very long times by a rapid quench 
to some final temperature $T$ which may be below or equal to the critical 
temperature $T_c$. The subsequent evolution then takes place at that fixed
temperature $T$. Main observables include the two-time autocorrelation function
$C(t,s)$ and the auto-response function $R(t,s)$ where $s$ is the waiting
time and $t$ the observation time. A convenient way to characterize the
distance of the system from equilibrium is through the 
fluctuation-dissipation ratio \cite{Cugl94a,Cugl94b,Bouc98}
\BEQ
X(t,s) = T R(t,s) \left( \frac{\partial C(t,s)}{\partial s} \right)^{-1}
\EEQ
At equilibrium, both $C$ and $R$ only depend on the time difference
$\tau=t-s$ and $X=X(\tau)=1$. Physically, this is realized if the initial
quench was made to some temperature $T>T_c$. Then for time scales large 
compared to the (finite) characteristic time scale $\tau_{\rm eq}$ 
the system relaxes exponentially fast towards equilibrium. 

On the other hand, if either $T<T_c$ or $T=T_c$, an infinite spin system
does not reach equilibrium on some finite time scale but instead undergoes
either phase ordering kinetics or non-equilibrium critical dynamics. In both
cases, two-time observables such as $C=C(t,s)$ and $R=R(t,s)$ depend
on {\em both} the waiting time $s$ and the observation time $t$ 
and not merely on their difference $\tau=t-s$. 
This breaking\footnote{This is associated with a
breaking of the fluctuation-dissipation theorem. The time scale on which this
breaking occurs has been studied very thouroughly in \cite{Zipp00}.} 
of time translation invariance is usally
referred to as {\em aging} and will be used in this sense from now on. 
Consequently, the fluctuation-dissipation ratio $X=X(t,s)\neq 1$ also becomes a 
non-trivial function of both $s$ and $t$. Empirically, it is well-established
\cite{Bray94} that the aging process is associated with dynamical scaling, 
that is in the scaling regime with times $1\ll s, \tau$ 
long enough after the initial quench one finds
\BEQ
C(t,s) \sim s^{-b} f_C(t/s) \;\; , \;\;
R(t,s) \sim s^{-1-a} f_R(t/s)
\EEQ 
where $a,b$ are non-equilibrium critical exponents and $f_C$ and $f_R$ are
scaling functions. Here and in the sequel we always have $t>s$. For large
arguments $x=t/s\gg 1$, these scaling functions typically behave as
\BEQ
f_{C}(x) \sim x^{-\lambda_C/z} \;\; , \;\;
f_{R}(x) \sim x^{-\lambda_R/z}
\EEQ
where $z$ is the dynamical exponent and $\lambda_C, \lambda_R$ are the
autocorrelation \cite{Fish88,Huse89} and autoresponse exponents.\footnote{The
values of the exponents $\lambda_C,\lambda_R$ (and also $a,b$) depend on
whether $T<T_c$ or $T=T_c$, but we shall use the same notation in both cases.}
Throughout this paper, we are only interested in this late-time regime where 
scaling occurs. 

Usually, the long-time behaviour after a quench from a completely
disordered state is studied. In this case, the available evidence as reviewed
in \cite{Godr01}, and based on results notably from the 
Glauber-Ising model \cite{Godr00a,Lipp00} and the spherical model with a 
non-conserved order parameter \cite{Cugl95,Godr00b} is consistent 
with the autocorrelation and autoreponse exponents being equal, 
$\lambda_C=\lambda_R$, and, for a quench exactly onto
criticality $T=T_c$, with the additional relation $a=b$. 
However, exceptions to this rule exist.
Consider the $2D$ XY model with a fully ordered zero-temperature initial
state and quenched `upwards' to some temperature $T<T_{\rm KT}$, where
$T_{\rm KT}$ is the Kosterlitz-Thouless transition temperature. 
A recent calculation by Berthier, Holdsworth and Sellitto \cite{Bert01} 
based on the spin-wave approximation 
has produced analytical results for both the autocorrelation and autoresponse
scaling functions $f_C(x)$ and $f_R(x)$ and found the scaling 
{\it ``in complete agreement with the
results obtained by Godr\`eche and Luck \cite{Godr00b} for the spherical model 
at the critical point''} \cite[p. 1809]{Bert01}. Indeed one has
$a=b=\eta/2$, where $\eta=\eta(T)$ is the usual static 
(and temperature-dependent) critical exponent, but the
autocorrelation exponent $\lambda_C/z=\eta/4$ and the autoresponse exponent
$\lambda_R/z=1+\eta/4$ \cite{Bert01} are different from each other.
We therefore ask ourselves what may generically
become of the relation between $\lambda_C$ and $\lambda_R$ for more 
general initial states than fully disordered or fully ordered ones. 
Beyond a case study in a given model, this allows a test of the generic
scaling and universality properties as reviewed abvoe and should provide
useful insight. 
The r\^ole of the initial conditions on aging in spin-glasses has been and
continues to be actively studied, see \cite{Cugl95,Zote02} and references 
therein. 

Questions of this sort are best addressed first in some non-trivial 
exactly solvable model
before being studied further through simulations. We shall therefore examine
for the kinetic spherical model described by a non-conserved Langevin equation 
(see section 2 for precise definitions) the r\^ole of long-range
correlations in the initial state which are characterized by a power law
for the Fourier transformation $\wit{C}(\vec{q})$ of the spin-spin
correlator, in the low-momentum limit $|\vec{q}|\to 0$
\BEQ
\wit{C}(\vec{q}) \sim |\vec{q}|^{\alpha}
\EEQ
and we shall study how the above scaling forms are affected, if at all,
by varying $\alpha$. We can therefore interpolate between a disordered initial
state with $\alpha=0$ and a fully ordered state ($\alpha=-d$) as considered
in \cite{Bert01}. While much is already known for $T<T_c$ \cite{Bray91}, 
we shall see that for $T=T_c$ there occur several new regimes for the long-time 
aging behaviour which depend on the value of $\alpha$. Our calculations
suggest the new scaling relation $\lambda_C =\lambda_R+\alpha_{\rm eff}$ where
$\alpha_{\rm eff}=\alpha_{\rm eff}(\alpha)$ is the effective value of $\alpha$
which actually describes the long-time behaviour. We shall show
that this also explains the results obtained in the $2D$ XY model \cite{Bert01}
and referred to above. 

Furthermore, introducing a new parameter into the kinetics permits an 
instructive test of the assertion by Godr\`eche and Luck \cite{Godr00b} 
that the limit fluctuation-dissipation ratio
\BEQ
X_{\infty} = \lim_{s\to\infty} \lim_{t\to\infty} X(t,s)
\EEQ
is a universal number and we shall indeed confirm this universality (see
section 3). 

Finally, we recall that recently the precise form of the autoresponse
function $R(t,s)$ was derived from conformal invariance \cite{Henk99}, 
for $t,s\gg1 $ being inside the scaling regime \cite{Henk01,Henk02}
\BEQ \label{1:ConfR}
R(t,s) = r_0 \left( \frac{t}{s}\right)^{1+a-\lambda_R/z} 
\left( t - s\right)^{-1-a}
\EEQ
where $r_0$ is a normalization constant. As we shall see in section 4, 
this functional
form is indeed recovered, but the exponents $a$ and $\lambda_R$ will
be functions of $\alpha$. 

The structure of this paper is as follows. In the next section, to give a
self-contained presentation, we recall the 
main steps of the general formalism for the exact calculation of
correlation and response functions. In section 3, we analyse these in the
scaling limit where aging occurs and find the scaling functions $f_{C}(x)$ and
$f_{R}(x)$ as functions of the initial conditions characterized by the
parameter $\alpha$. In section 4, the physical conclusions from these 
calculations are drawn. In the appendix, we shall briefly present
some results on initial conditions in the $1D$ Glauber-Ising model such that
the mean magnetization is non-vanishing.

\section{Formalism}

We now describe the calculation of the two-time autocorrelation and
autoresponse functions in the exactly solvable spherical model in $d$
spatial dimensions, for arbitrary initial conditions. 
Our calculation follows closely the standard lines established several times
in the past, see \cite{Bray91,Jans89,Newm90,Kiss93,Coni94,Cala02} 
for continuum field 
theories and \cite{Cugl95,Zipp00,Godr00b,Cann01,Corb02} for lattice models. 
The effects of specific initial conditions on
the long-time behaviour and aging will be analysed in the next section. 

The spherical model is defined in terms of real spin variables $S_{\vec{r}}$
attached to the sites of a $d$-dimensional hypercubic lattice and
subject to the constraint
\BEQ \label{2:SpCon}
\sum_{\vec{r}} S_{\vec{r}}^2 = {\cal N}
\EEQ
where $\cal N$ is the total number of sites, and the usual spin
Hamiltonian ${\cal H} = - \sum_{(\vec{r},\vec{r}')} S_{\vec{r}} S_{\vec{r}'}$
where the sum extends over nearest neighbour pairs only. The (non-conserved)
dynamics is given by the stochastic Langevin equation
\BEQ \label{2:eqMot}
\frac{\D S_{\vec{r}}}{\D t} = \sum_{\vec{s}(\vec{r})} S_{\vec{s}} -
(2d +\mathfrak{z}(t)) S_{\vec{r}} + \eta_{\vec{r}}(t)
\EEQ
where $\vec{s}(\vec{r})$ are the nearest neighbour sites of the site $\vec{r}$,
the Gaussian white noise $\eta_{\vec{r}}(t)$ has the correlation
\BEQ
\langle \eta_{\vec{r}}(t) \eta_{\vec{r}'}(t') \rangle = 
2T \delta_{\vec{r},\vec{r}'} \delta(t-t')
\EEQ
and $\mathfrak{z}(t)$ is determined by satisfying the spherical constraint 
({\ref{2:SpCon}) in the mean. By a Fourier transformation
\BEQ
\wit{f}(\vec{q}) = \sum_{\vec{r}} f_{\vec{r}} e^{-\II \vec{q}\cdot\vec{r}}
\;\; , \;\;
f_{\vec{r}} = (2\pi)^{-d} \int_{{\cal B}}\!\D\vec{q}\, \wit{f}(\vec{q}) 
e^{\II \vec{q}\cdot\vec{r}}
\EEQ 
where the integral is over the first Brillouin zone $\cal B$, 
eq.~(\ref{2:eqMot}) is transformed into
\BEQ
\frac{\partial \wit{S}(\vec{q},t)}{\partial t} = 
-\left[ \omega(\vec{q})+\mathfrak{z}(t)\right] 
\wit{S}(\vec{q},t) + \wit{\eta}(\vec{q},t)
\EEQ
where in addition, together with the $|\vec{q}|\to 0$ limit
\BEQ \label{2:omega-eta}
\omega(\vec{q})=2\sum_{i=1}^{d} \left( 1 -\cos q_i\right) \simeq q^2 \;\; ; \;\;
\langle \wit{\eta}(\vec{q},t) \wit{\eta}(\vec{q}',t')\rangle
=2T (2\pi)^d \delta^d(\vec{q}+\vec{q}')\delta(t-t')
\EEQ
The formal solution is \cite{Cugl95,Godr00b}
\BEQ
\wit{S}(\vec{q},t) = \frac{\exp(-\omega(\vec{q})t)}{\sqrt{g(T,t)\:}} 
\left[ \wit{S}(\vec{q},0) + \int_{0}^{t}\!\D t'\: 
e^{\omega(\vec{q})t'} \sqrt{g(T,t')\:}\: \wit{\eta}(\vec{q},t') \right] 
\;\; ; \;\;
g(T,t) = \exp\left( 2 \int_{0}^{t}\!\D t'\: \mathfrak{z}(t') \right) 
\EEQ
which forms the basis for all subsequent calculations. 

The Lagrange multiplier $g(T,t)$ is determined from the spherical constraint 
and the initial conditions. To see this, consider the equal-time 
spin-spin correlation function
\BEQ
C_{\vec{r}-\vec{r}'}(t) = \langle S_{\vec{r}}(t) S_{\vec{r}'}(t) \rangle
\EEQ
where spatial translation invariance is already taken into account.
Here and in the sequel, the brackets denote the average over the ensemble of
the initial conditions and over the thermal histories, i.e. the realizations
of the noise $\eta_{\vec{r}}(t)$. 
The spherical constraint (\ref{2:SpCon}) implies that
\BEQ \label{2:SpCon2}
C_{\vec{0}}(t) = \langle S_{\vec{r}}(t) S_{\vec{r}}(t) \rangle = 1
\EEQ
In Fourier space the equal-time correlator $\wit{C}(\vec{q},t)$ is obtained 
from
\BEQ
\langle \wit{S}(\vec{q},t)\wit{S}(\vec{q}',t)\rangle =
(2\pi)^{d} \delta^d(\vec{q}+\vec{q}') \wit{C}(\vec{q},t)
\EEQ
and is given by
\BEQ \label{2:Ctt}
\wit{C}(\vec{q},t) = \frac{\exp(-2\omega(\vec{q})t)}{g(T,t)} 
\left[ \wit{C}(\vec{q},0) + 2T \int_{0}^{t}\!\D t'\: 
e^{2\omega(\vec{q})t'} g(T,t') \right] 
\EEQ
From the spherical constraint (\ref{2:SpCon2}) we have in Fourier space that
$\int\!\D \vec{q}\, (2\pi)^{-d} \wit{C}(\vec{q},t) =1$ and this leads to the
following Volterra integral equation for $g(T,t)$, 
see \cite{Cugl95,Godr00b}
\BEQ \label{2:Volt}
g(T,t) = A(t) + 2T \int_{0}^{t}\!\D t'\, f(t-t') g(T,t')
\EEQ
where the two auxiliary functions $f(t)$ and $A(t)$ are defined as follows
\BEQ
f(t) = \frac{1}{(2\pi)^d} \int\!\D \vec{q}\, e^{-2\omega(\vec{q})t} 
= \left( e^{-4t} I_0(4t)\right)^d \;\; , \;\;
A(t) = \frac{1}{(2\pi)^d} \int\!\D \vec{q}\, e^{-2\omega(\vec{q})t} 
\wit{C}(\vec{q},0)
\EEQ
and where $I_0$ is a modified Bessel function \cite{Abra65}. 
The solution of equation (\ref{2:Volt}) is found from a Laplace transformation
\BEQ \label{2:fA}
\lap{f}(p) = \int_{0}^{\infty} \!\D t\, f(t) e^{-pt}
\EEQ 
and is given by
\BEQ \label{2:Lapg}
\lap{g}(T,p) = \frac{\lap{A}(p)}{1-2T\: \lap{f}(p)}
\EEQ

Therefore the entire evolution of the system, starting from a freely chosen 
initial condition, can be described in terms of the properties of the functions
$f(t)$ and $A(t)$. In particular, the initial state is characterized 
exclusively by the initial correlator $\wit{C}(\vec{q},0)$ and these data 
enter explicitly only into the function $A(t)$. 
The simplest case may be considered to be an initial state without any
correlations. Then $C_{\vec{r}}(0)=\delta_{\vec{r},\vec{0}}$, therefore
$\wit{C}(\vec{q},0)=1$ and thus $A(t)=f(t)$. This case has been
analysed in great detail, see \cite{Cugl95,Zipp00,Corb02} and in particular
\cite{Godr00b}. 

We postpone the analysis of the effects of initial conditions given by
$\wit{C}(\vec{q},0)$ to the next section
and now give the general expressions for the two-time correlators and the
two-time response functions. The two-time correlation function is
defined as
\BEQ
C_{\vec{r}-\vec{r}'}(t,s) = \langle S_{\vec{r}}(t) S_{\vec{r}'}(s)\rangle
\EEQ
where $s$ is the waiting time, $t$ the observation time and $t\geq s\geq 0$
always. In Fourier space, it is easy to see that
\BEQ
\wit{C}(\vec{q},t,s) = \wit{C}(\vec{q},s) e^{-\omega(\vec{q})(t-s)}
\sqrt{\frac{g(T,s)}{g(T,t)}}
\EEQ
and where the expression (\ref{2:Ctt}) for the single-time correlator has been
used. Below, we shall be mainly be interested in the two-time
autocorrelation 
\BEA
C(t,s) &=& C_{\vec{0}}(t,s) = 
(2\pi)^{-d} \int\!\D\vec{q}\, \wit{C}(\vec{q},t,s) 
\nonumber \\
&=& \frac{1}{\sqrt{g(T,t)g(T,s)}} 
\left[ A\left(\frac{t+s}{2}\right) + 2T \int_{0}^{s}\!\D s'\: 
f\left(\frac{t+s}{2}-s'\right) g\left(T,s'\right) \right]
\label{2:Cts}
\EEA
which we shall analyse in the next section. The response function
is obtained in the usual way \cite{Newm90,Kiss93,Cugl95,Godr00b} 
by adding a small magnetic
field term $\delta{\cal H} = - \sum_{\vec{r}} h_{\vec{r}}(t) S_{\vec{r}}(t)$
to the Hamiltonian. This leads to an extra term $h_{\vec{r}}(t)$ on the
right-hand side of the Langevin equation (\ref{2:eqMot}). Provided that
causality and spatial translation invariance hold, we have to first order
\BEQ
\langle S_{\vec{r}}(t)\rangle = \int_{0}^{t}\!\D s\,
\sum_{\vec{r}'} R_{\vec{r}-\vec{r}'}(t,s) h_{\vec{r}'}(s) +\ldots
\EEQ
which defines the (linear) response function $R_{\vec{r}}(t,s)$. The
calculation is completely standard \cite{Newm90,Cugl95,Godr00b}
and we merely quote the result
\BEQ \label{2:Rqts}
\wit{R}(\vec{q},t,s) = \left. 
\frac{\delta\langle\wit{S}(\vec{q},t)\rangle}{\delta \wit{h}(\vec{q},s)}
\right|_{h_{\vec{r}}=0}
= e^{-\omega(\vec{q})(t-s)} \sqrt{\frac{g(T,s)}{g(T,t)}}
\EEQ
Again, we shall be mainly concerned with the autoresponse function
\BEQ \label{2:Rts}
R(t,s) = R_{\vec{0}}(t,s) = (2\pi)^{-d} \int\!\D\vec{q}\, \wit{R}(\vec{q},t,s)
= f\left(\frac{t-s}{2}\right) \sqrt{\frac{g(T,s)}{g(T,t)}}
\EEQ

Eqs.~(\ref{2:Ctt},\ref{2:Cts},\ref{2:Rts}) give the single-time
correlation function and the two-time autocorrelation and autoresponse
functions, respectively, for yet arbitrary initial conditions. 
Together with the spherical constraint 
(\ref{2:Volt}), which fixes the function $g(T,t)$ in terms of the
auxiliary functions $f(t)$ and $A(t)$, these are the main results of this 
section. 

\section{Analysis}

It is our aim to consider the effects of long-range correlations in the initial
state on the long-time and aging behaviour of the model. For the long-time
behaviour, only the low-momentum regime should be relevant, which we
take to be of the form \cite{Bray91}
\BEQ \label{3:Ini}
\wit{C}(\vec{q},0)= c_{0} + c_{\alpha} |\vec{q}|^{\alpha}
\EEQ 
where $\alpha$ is a free parameter and $c_0,c_{\alpha}$ are normalization 
constants. For $\alpha>0$, the long-time behaviour depends to leading order
only on the first term while for $\alpha<0$, we have effetively $c_0=0$,
to leading order. For the purposes of this paper, namely the study of the
r\^ole of long-range correlations in the initial conditions as parametrized
by $\alpha$, it is sufficient to study the case $c_0=0$, which we shall assume
from now on. For $\alpha<0$, this corresponds to initial correlations of the 
form $C_{\vec{r}}(0)\sim r^{-d-\alpha}$ for large distances $r=|\vec{r}|$. 
This initial condition should lead to the following $t\to\infty$ 
asymptotics of the auxiliary function 
\BEQ \label{3:Aa}
A(t) \simeq a_{\alpha} t^{-(d+\alpha)/2} \;\; , \;\; 
a_{\alpha} = c_{\alpha} 
(2\pi)^{-d} \int\!\D \vec{u} \: e^{-2\vec{u}^2} u^{\alpha} 
\EEQ
The case without initial correlations
considered previously \cite{Cugl95,Zipp00,Godr00b,Corb02} corresponds to 
$\wit{C}(\vec{q},0)=1$ and is recovered if we formally set
$\alpha=0$. Any differences between the scaling behaviour coming from the
initial conditions (\ref{3:Ini}) and for those analysed previously must 
related to a different long-time behaviour of the functions $A(t)$ and $f(t)$
(it is well known that for $t$ large, one has $f(t)\simeq (8\pi t)^{-d/2}$). 
It is the purpose of this section to carry through the technical analysis
and especially to obtain the explicit scaling functions for the autocorrelation,
autoresponse and the fluctuation-dissipation ratio. The physical discussion
of the results will be presented in section 4.  

The single-time correlator $C_{\vec{r}}(t)$ has been analysed in great detail 
by Coniglio, Ruggiero and Zanetti \cite{Coni94} in the context of 
coarse-grained field theory. 

We now analyse the long-time behaviour of the two-time correlators and
response functions. For the questions we are interested in, namely
the breaking of the fluctuation-dissipation theorem and/or aging
effects, it is enough to restrict to temperatures at or below 
criticality. Indeed, for $T>T_c$, the system will simply relax to
equilibrium within a finite time scale $\tau_{\rm eq}\sim (T-T_c)^{-2\nu}$,
where $\nu$ is the known equilibrium correlation length exponent, see
\cite{Godr00b} and references therein. For the
same reason, we restrict to $d>2$ throughout such that there is always a
phase transition at a non-vanishing critical temperature $T_c>0$.  
In addition, the early stages of the evolution of the system (notably the
evolution of the fluctuation-dissipation ratio) will depend
on whether the system was initially prepared in an equilibrium state or not
but this information requires more knowledge on the initial state
than merely its low-momentum behaviour eq.~(\ref{3:Ini}). 
Rather, the behaviour we are interested in is described by
the scaling or aging
limit which is reached when both the waiting time $s$ and the observation
time $t$ become simultaneously large, that is $1\ll s \sim t-s$. 
As we shall be interested in the scaling properties of
$C(t,s)$ and $R(t,s)$, it is useful to work with the dimensionless scaling 
variable
\BEQ
x = \frac{t}{s} > 1
\EEQ

Taking the scaling limit $t,s\to\infty$ with $x$ fixed, 
it is convenient to rewrite the results 
(\ref{2:Cts},\ref{2:Rts}) of the previous section as follows
\BEA
C(t,s) &\simeq& \frac{s^{-d/2+1}}{\sqrt{g(T,s)g(T,t)}} 
\left[ a_{\alpha} s^{-\alpha/2-1} 
\left(\frac{x+1}{2}\right)^{-\frac{1}{2}(d+\alpha)}
+\frac{2T}{(4\pi)^{d/2}} \int_{0}^{1} \!\D\theta\, g(T,s\theta) 
\left(x+1-2\theta\right)^{-d/2} \right] 
\nonumber \\
R(t,s) &\simeq& (4\pi)^{-d/2} s^{-d/2} \left(x-1\right)^{-d/2}
\sqrt{\frac{g(T,s)}{g(T,t)}}
\label{3:CRtsSkal}
\EEA
where we have neglected correction terms to the leading scaling behaviour.
Next, we need the leading behaviour of $g(T,t)$ for $t$ large. 

\subsection{Non-equilibrium critical dynamics} 
We first consider the case when $T=T_c$, the equilibrium critical temperature. 
In general, we expect the leading asymptotics of $g(T,t)$ to be of the
form
\BEQ \label{3:gtlong}
g(T,t) \simeq g_0(\alpha,d) t^{\psi(\alpha,d)} \;\; ; \;\; t\gg 1
\EEQ
where the dependence of $\psi=\psi(\alpha,d)$ must be calculated and 
$g_0$ is a constant. We determine $\psi$ by finding first from
eq.~(\ref{2:Lapg}) the low-$p$ behaviour of $\lap{g}(T,p)$, which in turn
is given by the small-$p$
behaviours of $\lap{f}(p)$ and $\lap{A}(p)$. These may be found from the
integral representations (\ref{2:fA}) using well-established techniques
\cite{Mont65,Godr00b}. At the critical point, the results are as follows: 
\BEA
\lap{f}(p) &\simeq & A_1 + \frac{\Gamma(1-d/2)}{(8\pi)^{d/2}} 
p^{d/2-1} + \ldots \;\; ; \;\; 2 < d < 4 
\nonumber \\
\lap{f}(p) &\simeq & A_1 - A_2 p+ \ldots \hspace{2.4truecm} \;\; ; \;\; d > 4
\EEA
where $T_c = 1/(2A_1)$ is the equilibrium critical temperature and 
$A_k=(2\pi)^{-d}\int\!\D\vec{q}\,(2\omega(\vec{q}))^{-k}$,
$k=1,2,\ldots$ which exist for $d>2k$. Similarly, we find
\BEA
\lap{A}(p) &\simeq & a_{\alpha}\Gamma\left(1-\frac{1}{2}(d+\alpha)\right)\: 
p^{-1+(d+\alpha)/2} 
\;\; ; \;\; 0 < d+\alpha < 2 
\nonumber \\
\lap{A}(p) &\simeq & B_1 + O\left(p, p^{-1+(d+\alpha)/2}\right)
\hspace{1.6truecm} \;\; ; \;\; d+\alpha > 2
\EEA
where $a_{\alpha}$ is given in eq.~(\ref{3:Aa}) and
$B_1 = (2\pi)^{-d}\int\!\D\vec{q}\,\wit{C}(\vec{q},0) (2\omega(\vec{q}))^{-1}$
which exists for $d+\alpha>2$ (if either $d=2,4$ or $d+\alpha=0,2$ additional
logarithmic factors will be present, which we shall discard throughout). 
From this, $\lap{g}(T,p)$ follows and transforming back to $g(T,t)$, the
exponent $\psi(\alpha,d)$ describing the leading behaviour for $t$ large is
found. We collect the results in table~\ref{tab1} and identify five regimes
where the behaviour of $g(T,t)$ is different. It is convenient to characterize
these regimes in terms of an effective dimension
\BEQ \label{3:D}
D = d+\alpha + 2
\EEQ
the meaning of which we shall discuss in section 4. 
\begin{table}
\caption{Values of the exponent $\psi=\psi(\alpha,d)$ which describes the
long-time behaviour of $g(T_c,t)$ according to 
eq.~(\ref{3:gtlong}) and the exponent $\vph=1+\psi+\alpha/2$ 
in the five different scaling regimes at criticality.\label{tab1}}
\begin{center} \begin{tabular}{|c|ccc|cc|} \hline
Regime & \multicolumn{3}{c|}{conditions} & $\psi$ & $\vph$ \\ \hline
I   & $2<d<4$ & $2<D<4$ & & $-1-\alpha/2$     & $0$ \\
II  & $4<d$   & $2<D<4$ & & $1-(d+\alpha)/2$  & $(4-d)/2$ \\
III & $2<d<4$ & $4<D$   & & $d/2-2$           & $(d+\alpha)/2-1$ \\
IV  & $4<d$   & $4<D$   & $\alpha>-2$ & $0$   & $\alpha/2+1$ \\ 
V   & $4<d$   & $4<D$   & $\alpha<-2$ & $0$   & $\alpha/2+1$ \\ \hline
\end{tabular}\end{center}
\end{table} 
From these results, we find the following scaling forms 
\BEA
C(t,s) &=& (4\pi)^{-d/2} s^{-d/2+1} \left[ 
s^{-\vph} M_0(x) + K_0(x) \right]
\nonumber \\
R(t,s) &=& (4\pi)^{-d/2} s^{-d/2} (x-1)^{-d/2} x^{-\psi/2}
\EEA
where the values of the exponent $\vph=1+\psi+\alpha/2$ are also listed in 
table~\ref{tab1} and
\newpage
\BEA
M_0(x) &=& \frac{a_{\alpha} (4\pi)^{d/2}}{g_0} x^{-\psi/2} 
\left(\frac{x+1}{2}\right)^{-(d+\alpha)/2}
\nonumber \\
K_0(x) &=& 2T x^{-\psi/2} \int_{0}^{1} \!\D w\, w^{\psi} (x+1-2w)^{-d/2}
\EEA
Furthermore, the scaling of the fluctuation-dissipation ratio can be
written in the form
\BEQ
X(t,s) = T \frac{ (x-1)^{-d/2} x^{-\psi/2}}{s^{-\vph} M(x) + K(x)}
\EEQ
where the functions $M(x)$ and $K(x)$ are defined by
\BEA
M(x) &=& -\left(\frac{d+\alpha}{2}+\psi\right)M_0(x)-x\frac{\D M_0(x)}{\D x}
\nonumber \\
K(x) &=& \left(1-\frac{d}{2}\right) K_0(x) - x \frac{\D K_0(x)}{\D x}
\EEA
and we see that the scaling of $C(t,s)$ and consequently of $X(t,s)$ 
depends on the sign of $\vph$. For that reason, the regimes IV and V
have to be distinguished. 

We can now list the scaling functions for both the autocorrelation and 
autoresponse functions
\BEQ \label{3:fcfr}
C(t,s) = (4\pi)^{-d/2} T_c\: s^{-b} f_C(x) \;\; , \;\;
R(t,s) = (4\pi)^{-d/2} s^{-1-a} f_R(x)
\EEQ
together with the fluctuation-dissipation ratio. The results for the
exponents $a$ and $b$ are given in table~\ref{tab2} below. For the response
function, the scaling function simply is in all five regimes
\BEQ \label{3:fRskal}
f_{R}(x) =  (x-1)^{-d/2} x^{-\psi/2}
\EEQ
and the values of $\psi$ can be read off from table~\ref{tab1}. 
We shall return to this result in section 4. 

The functions $f_C(x)$ and $X(t,s)$ are listed below: for regime I, we find
\BEA
f_C(x) &=& \left[ 2^{1+\alpha/2} 
\left|\frac{\Gamma(1-d/2)\Gamma(-\alpha/2)}{\Gamma(1-(d+\alpha)/2)}\right|
x^{-(d+\alpha/2-1)/2} +2 x^{1/2+\alpha/4} \int_{0}^{1}\!\D y\,
y^{-1-\alpha/2} (x+1-2y)^{-d/2} \right] \nonumber \\
&\simeq&  2^{1+\alpha/2} 
\left|\frac{\Gamma(1-d/2)\Gamma(-\alpha/2)}{\Gamma(1-(d+\alpha)/2)}\right|
x^{-d/2-\alpha/4+1/2} \;\; , \;\; x\to\infty
\EEA
The fluctuation dissipation ratio becomes in the scaling limit a function
of $x$ only, which reads for $\alpha\ne -2$
\BEA
X = X(x) &=& (x-1)^{-d/2} 
\left[ -\left|\frac{\Gamma(1-d/2)\Gamma(-\alpha/2)}
{\Gamma(1-(d+\alpha)/2)}\right|  2^{\alpha/2} \left(\frac{d+\alpha}{x+1}-
\frac{\alpha}{2}-1\right) x^{-(d+\alpha)/2} \right. \nonumber \\
& & \left. + 2 
\int_{0}^{1}\!\D y\, y^{-1-\alpha/2} (x+1-2y)^{-d/2}
\left(\frac{1}{2}-\frac{\alpha}{4}+\frac{d}{2}\frac{2y-1}{x+1-2y}\right)
\right]^{-1} \nonumber \\
&\simeq& \left|\frac{\Gamma(1-d/2)\Gamma(-\alpha/2)}
{\Gamma(1-(d+\alpha)/2)}\right| \frac{2}{2+\alpha} (2x)^{\alpha/2}
\;\; , \;\; x\to\infty
\EEA
together with the leading behaviour for infinitely separated timescales
$x\to\infty$. For $\alpha=-2$, we find
\BEQ
X = X(x) = \left[ 1+\left(\frac{x-1}{x+1}\right)^{d/2}
\left(1-\left(\frac{x}{x+1}\right)^{1-d/2}\right)\right]^{-1}
\simeq 1 \;\; , \;\; x\to\infty
\EEQ
Therefore, if $\alpha\ne -2$, the limit fluctuation-dissipation ratio
has the universal value $X_{\infty}=0$ but the approach towards that limit
does depend on the initial condition, while for $\alpha=-2$, we have 
$X_{\infty}=1$ signalling that an equilibrium state will be reached. 

\begin{figure}[ht]
\centerline{\epsfxsize=4.75in\epsfbox
{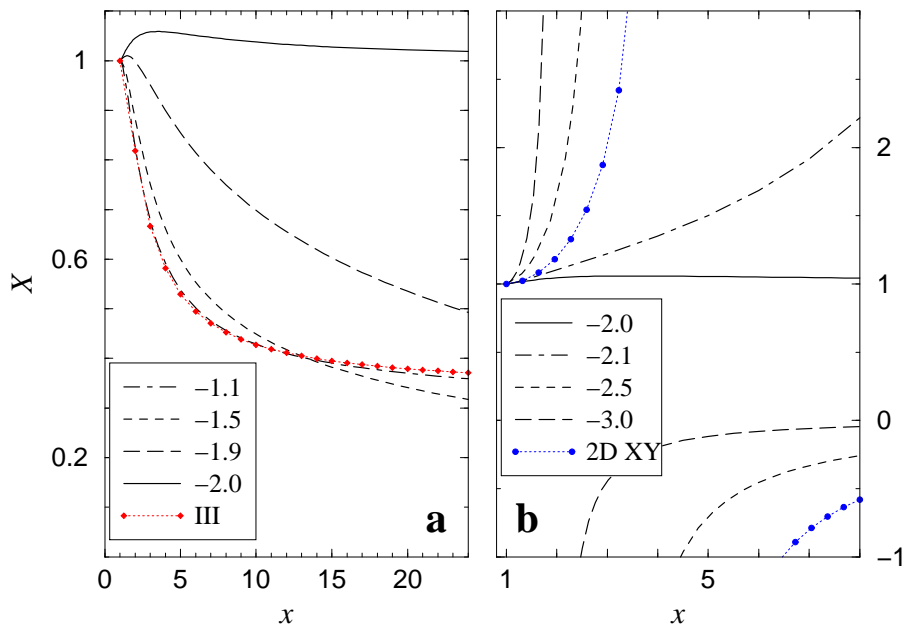}}
\caption{The scaling of the fluctuation-dissipation ratio $X(x)$ as a function
of $x=t/s$ in regime I for $d=3$ and for several values of $\alpha$. In panel
a, we also include the result $X^{(III)}(x)$ found for regime III and given in
eq.~(\ref{3:XIII}). In panel b we also show the function 
$X_{\rm 2D\:{\rm XY}}(x)$ obtained in the $2D$ XY model in the spin-wave 
approximation \protect{\cite{Bert01}} and given in eq.~(\ref{4:XY2D}). 
\label{Abb2}}
\end{figure}
The behaviour of the scaling function $X(x)$ is illustrated in 
figure~\ref{Abb2}. In the left part (figure~\ref{Abb2}a) we consider 
initial states with $\alpha\geq -2$ which are more disordered than the
critical equilibrium state. In all cases, one starts from equilibrium at equal
times, since $X(1)=1$. If $\alpha\ne -2$, the fluctuation-dissipation ratio
decays with $x$ increasing. However, if $\alpha$ approaches the border between 
regions I and region III, the scaling function $X^{(I)}(x)$ obtained for the 
region I goes over smoothly into the one found for region III. Close to that
boundary, quite large values of $x$ are needed in order to distinguish these
functions. On the other hand, for 
$\alpha=-2$, the long-range properties of the initial state are the same as
for the equilibrium state. Although the system departs initially from
equilibrium, since $X(x)>1$, the non-equilibrium short-range correlations
are succesively equilibrated and the system finally arrives at the
equilibrium value of the limit fluctuation-dissipation ratio $X_{\infty}=1$. 
If we now consider the case $\alpha\leq -2$ (figure~\ref{Abb2}b) with an
initial state more ordered than the equilibrium state, the behaviour is 
completely different. Starting from $X(1)=1$, the fluctuation-dissipation 
ratio increases with $x$ and encounters at some finite value 
$x_s=x_s(d,\alpha)$ a singularity. For $x>x_s$, it is negative and rapidly
decays to zero with increasing $x$.  

For regime II, we find
\BEA
f_C(x) &=& 2(4\pi)^{d/2} A_2 \left(1-\frac{d+\alpha}{2}\right) 
x^{(d+\alpha)/4-1/2} 
\left(\frac{x+1}{2}\right)^{-(d+\alpha)/2}  \nonumber \\
X(xs,s)  &=& \frac{-(4\pi)^{-d/2}(x-1)^{-d/2}}{A_2 (1-(d+\alpha)/2)
(1+(d+\alpha)(3x+1)(2x+2)^{-1})} 
\left(\frac{x+1}{2}\right)^{(d+\alpha)/2}\: s^{-(d-4)/2} 
\EEA
where the integral $A_2$ was defined in section 3. 
Therefore for any value of $x$, $X(xs,s)=0$ in the scaling limit $s\to\infty$, 
since $d>4$ here. 

For regime III, we have
\BEA
f_C(x) &=& \frac{4}{d-2}\frac{(x-1)^{-d/2+1} x^{1-d/4}}{x+1} \nonumber \\
X = X(x) &=& \left(1+\frac{2}{d-2}\left(\frac{x-1}{x+1}\right)^{2}\right)^{-1}
\label{3:XIII}
\EEA
which reproduces the scaling functions previously found \cite{Godr00b} for
$\alpha=0$. The entire scaling functions turn out to
be completely independent of $\alpha$. Therefore, the scaling behaviour
in this region is governed by the effective value $\alpha_{\rm eff}=0$ 
in the initial condition (\ref{3:Ini}) 
and all scaling functions are universal. In 
particular, the limit fluctuation-dissipation ratio 
$X_{\infty}=\lim_{x\to\infty} X(x)=1-2/d$ is a universal number, 
as expected.  

We have for the case of regime IV
\BEA
f_C(x) &=& \frac{2}{d-2} \left((x-1)^{-d/2+1}-(x+1)^{-d/2+1}\right) \nonumber\\
X = X(x) &=& \left( 1+\left(\frac{x-1}{x+1}\right)^{d/2}\right)^{-1} 
\EEA
which again agrees completely with the earlier results found for $\alpha=0$ 
\cite{Godr00b}. Again, the scaling functions are independent of $\alpha$
and thus $\alpha_{\rm eff}=0$. In particular $X_{\infty}=1/2$ is a 
universal number.

Finally, in the regime V, we find  
\BEA
f_C(x) &=& 2(4\pi)^{d/2} A_2 B_V 
\left(\frac{x+1}{2}\right)^{-(d+\alpha)/2}
\nonumber \\
X(xs,s) &=& \frac{-(4\pi)^{-d/2} (x-1)^{-d/2} (x+1)}{A_2 B_V (d+\alpha) (2x+1)}
\left(\frac{x+1}{2}\right)^{(d+\alpha)/2} \:s^{1+\alpha/2} 
\EEA
where the constant $B_V$ is given by
\BEQ
B_V = \left[ \int\!\D \vec{u}\, u^{\alpha} e^{-2 \vec{u}^2} \right] \cdot
\left[ \int\!\D \vec{u}\, \frac{u^{\alpha}}{2\omega(\vec{u})} \right]^{-1}
\EEQ
Therefore $X(xs,s)=0$ in the scaling limit, since $1+\alpha/2<0$. 

{}From the $x\to\infty$ asymptotics
\BEQ \label{3:fasymp}
f_{C}(x) \sim x^{-\lambda_C/z} \;\; , \;\;
f_{R}(x) \sim x^{-\lambda_R/z}
\EEQ
and the known fact that $z=2$ \cite{Bray94} 
for the spherical model at and below $T_c$,
we read off the critical autocorrelation and autoresponse exponents 
$\lambda_C$ and $\lambda_R$ and collect the results in table~\ref{tab2}. 
They will be discussed in section 4.   
\begin{table}
\caption{Values of the critical autocorrelation and autoresponse exponents 
$a$, $b$, $\lambda_C$ and $\lambda_R$ as defined in 
eqs. (\ref{3:fcfr},\ref{3:fasymp}) in the five scaling regimes.\label{tab2}}
\begin{center}
\begin{tabular}{|c|cccc|} \hline
Regime & $a$ & $b$ & $\lambda_C$ & $\lambda_R$ \\ \hline
I   & $d/2-1$ & $d/2-1$        & $d+\alpha/2-1$   & $d-\alpha/2-1$   \\
II  & $d/2-1$ & $1$            & $(d+\alpha)/2+1$ & $(d-\alpha)/2+1$ \\
III & $d/2-1$ & $d/2-1$        & $3d/2-2$         & $3d/2-2$         \\
IV  & $d/2-1$ & $d/2-1$        & $d$              & $d$              \\
V   & $d/2-1$ & $d/2+\alpha/2$ & $d+\alpha$       & $d$              \\ \hline
\end{tabular} \end{center}
\end{table}

\subsection{Phase ordering} 

Having found the scaling behaviour at criticality, we now turn to the
ordered phase, where $T<T_c$. This case was considered long ago in the context
of the coarse-grained O($n$)-symmetric field theory in the limit
$n\to\infty$ \cite{Bray91,Newm90}. The relations (\ref{3:CRtsSkal}) remain
valid, but the Lagrange multiplier $g(T,t)$ now has for large times
the leading behaviour
\BEQ
g(T,t) \simeq \frac{a_{\alpha}}{M_{\rm eq}^{2}} t^{-(d+\alpha)/2}
\EEQ
where $M_{\rm eq}^{2}=1-T/T_c$ is the squared equilibrium magnetization. 
In the scaling limit $s\to\infty$, $t\to\infty$ but $x=t/s$ fixed, one has
for $\alpha\leq 0$
\BEA
C(t,s) &=& M_{\rm eq}^{2} \left(\frac{(x+1)^{2}}{4x}\right)^{-(d+\alpha)/4}
\nonumber \\
R(t,s) &=& (4\pi)^{-d/2} s^{-d/2} (x-1)^{-d/2} x^{(d+\alpha)/4} 
\label{3:CRXbas}\\
X(t,s) &=& \frac{4^{1-(d+\alpha)/4} T}{(4\pi)^{d/2} (d+\alpha) M_{\rm eq}^{2}}
\: \frac{(x+1)^{d/2+\alpha+1}}{(x-1)^{d/2+1}} \: s^{-d/2+1}
\nonumber
\EEA
For $\alpha=0$, these results were already known \cite{Newm90,Godr00b,Corb02}
(for $\alpha>0$, the constant term in (\ref{3:Ini}) is dominating the long-time
behaviour). As expected, the fluctuation-dissipation ratio $X=0$ in the scaling
limit. The scaling functions $f_C(x)$ and $f_R(x)$ are defined as
\BEQ 
C(t,s) =  M_{\rm eq}^{2} f_C(x) \;\; , \;\;
R(t,s) = (4\pi)^{-d/2} s^{-1-a} f_R(x)
\EEQ
and we obtain 
\BEQ \label{3:Tbas1}
a = \frac{d}{2}-1 \;\; , \;\; 
f_C(x) = \left(\frac{(x+1)^{2}}{4x}\right)^{-(d+\alpha)/4} \;\; , \;\;
f_R(x) = (x-1)^{-d/2} x^{(d+\alpha)/4}
\EEQ
and formally $b=0$ if we were to compare with the scaling (\ref{3:fcfr}) at 
criticality. From these, using again eq. (\ref{3:fasymp}), we have
the autocorrelation and autoresponse exponents
\BEQ \label{3:Tbas2}
\lambda_C = \frac{1}{2}(d+\alpha) \;\; , \;\;
\lambda_R = \frac{1}{2}(d-\alpha)
\EEQ
These results, valid for $T<T_c$, supplement those for the critical case
$T=T_c$ given in table~\ref{tab2}. For arbitrary $\alpha<0$ the value of 
$\lambda_C$ was already known \cite{Bray91}.\footnote{For the O($n$)-model
with $n$ {\em finite}, $\lambda_C=d/2$ for $0>\alpha>\alpha_c$, where 
$\alpha_c$ is known to leading order in $1/n$. The result of eq.~(\ref{3:Tbas2})
for $\lambda_C$ only holds true if $\alpha<\alpha_c$. In the
$n\to\infty$ limit, $\alpha_c=0$ \cite{Bray91}.}
A long time ago, Newman and Bray \cite{Newm90} studied the case of zero 
waiting time $s=0$. {}From eqs (\ref{2:Cts},\ref{2:Rts}) and the 
initial condition $g(T,0)=1$, one has
\BEQ
C(t,0) = \frac{A(t/2)}{\sqrt{g(T,t)}} \sim {M_{\rm eq}} 
t^{-(d+\alpha)/4}
\;\; , \;\;
R(t,0) = \frac{f(t/2)}{\sqrt{g(T,t)}} \sim {M_{\rm eq}}
t^{-(d-\alpha)/4}
\EEQ
in full agreement with their results.

\section{Discussion}

The long-time behaviour of the kinetic spherical model depends on the spatial
dimension $d$ and the parameter $\alpha$ which charcaterizes the initial
condition eq. (\ref{3:Ini}). We have already introduced 
the effective dimension $D$, see
eq. (\ref{3:D}). In terms of these, the equilibrium correlation function
$C_{\rm eq}(\vec{r})$ at criticality and the initial correlation function 
$C_{\rm ini}(\vec{r})=C_{\vec{r}}(0)$ scale for large distances $r=|\vec{r}|$ as
\BEQ \label{4:EquiIni}
C_{\rm eq}(\vec{r}) \sim r^{-(d-2)} \;\; , \;\;
C_{\rm ini}(\vec{r}) \sim r^{-(D-2)}
\EEQ
(since the equilibrium critical exponent $\eta=0$ for the sperical model). 
Therefore, we have studied the influence of the fluctuations of an effectively 
$D$-dimensional system onto a model defined in $d$ 
dimensions.\footnote{Practically, long-range initial conditions as studied here 
might be realized by coupling the degrees of freedom of the model under study 
to those of another system {\em at} criticality. Alternatively, one might 
consider a spin system with long-range interactions $J=J(\vec{r})$ and use this
to prepare an initial state with long-range correlations.} We can conclude: 

\begin{enumerate}
\item
In the scaling limit which describes the aging of the system, the results
obtained for a quench to a temperature $T<T_c$ below the critical point
and given in eqs. (\ref{3:Tbas1},\ref{3:Tbas2}) show that the effect
of initial long-range correlations persists for all times of the ongoing
non-equilibrium behaviour, but without provoking any qualitative changes,
as observed long ago \cite{Newm90,Bray91}. Remarkably enough, however, for
times $1\ll t-s\ll s$ which in the scaling limit corresponds to $x\simeq 1$ 
and for any initial condition, the two-time
autocorrelations saturate at a plateau value $C(x=1)\simeq M_{\rm eq}^2$,
see eq.~(\ref{3:CRXbas}).
For larger differences between the waiting time $s$ and the observation
time $t$, $C(t,s)$ decays to zero according to a power law which now again
depends on the initial conditions and the systems shows thus a sort of
memory of its initial state.  

\begin{figure}[ht]
\centerline{\epsfxsize=3.75in\epsfbox
{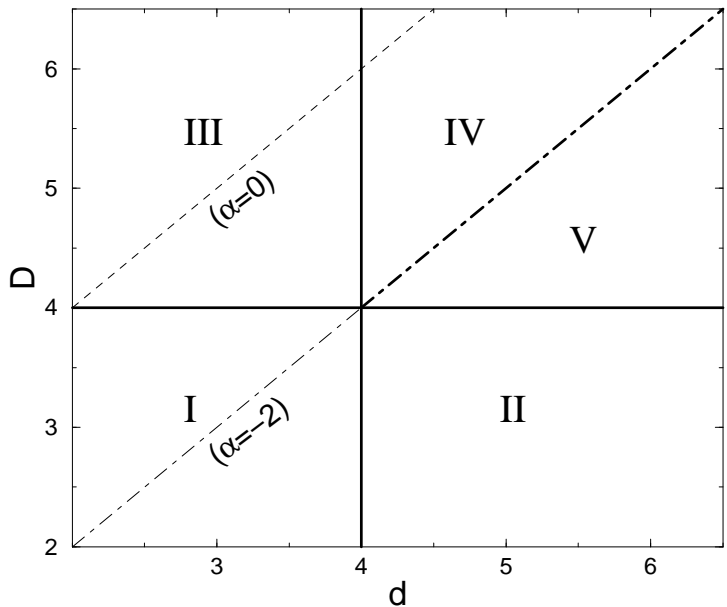}}
\caption{Kinetic phase diagram at $T=T_c$ depending on the spatial dimension
$d$ and on the effective dimension $D=d+\alpha+2$ where $\alpha$ is the
exponent of the initial condition eq.~(\ref{3:Ini}). 
The five different scaling regimes I, \ldots,
V are shown. The dashed line ($\alpha=0$) corresponds to the case
of a completely disordered initial state, the dash-dotted line ($\alpha=-2$)
corresponds to an equilibrium initial state at $T=T_c$ and the fully ordered
zero-temperature initial state corresponds to the line $D=2$.
\label{Abb1}}
\end{figure}
Surprisingly, the results for a quench
precisely to $T=T_c$ are qualitatively different. They are collected in
tables~\ref{tab1} and \ref{tab2} and the resulting phase diagram is shown 
in figure~\ref{Abb1}. Indeed, we find five regimes with different aging
behaviours. These regimes are distinguished by the presence (when $2<d<4$
and/or $2<D<4$) of strong critical fluctuations in either the equilibrium
or the initial state, respectively, or else their absence (when $d>4$
and/or $D>4$) when either the equilibrium or the initial state are in the
mean field regime. In addition, if both $d>4$ and $D>4$, it is of relevance
whether the initial correlations decay faster than those in the equilibrium
state or not. If $d=D$, the system is prepared in its critical equilibrium
state, the fluctuation-dissipation theorem is valid and no aging occurs. That
is the situation of {\em equilibrium} critical dynamics. Specifically,
we find
\begin{enumerate}
\item Only if the initial correlations are in the mean field regime, i.e.
$D>4$, and if in addition the initial correlations decay faster than in
equilibrium, the system may show a non-vanishing value of the 
fluctuation-dissipation ration $X_{\infty}$. In these cases (regimes III/IV), 
the entire scaling functions $f_C(x)$, $f_R(x)$ and $X(x)$ do {\em not\/} 
depend of the initial conditions at all and agree with the previously known
results obtained for a disordered initial state of infinite temperature, see
\cite{Cugl95,Godr00b,Corb02}. One therefore has effective initial conditions
such that $\alpha_{\rm eff}=0$. That finding is in full agreement with
the expected \cite{Godr00b,Godr01} universality of these scaling functions 
and of $X_{\infty}$ in particular. 
\item A non-trivial result for the scaling of the fluctuation-dissipation 
ratio $X(x)$ is found if both the equilibrium and the initial states are 
fluctuation-dominated (regime I). The limit value
for very large separations $x=t/s\to\infty$ between the waiting time $s$ and 
the observation time $t$ is the universal value $X_{\infty}=0$, provided
that $\alpha\ne -2$, but the 
approach towards this limit depends on the initial condition through
$X(x)\sim x^{-|\alpha|/2}$, as was illustrated in figure~\ref{Abb2}. 
It is not impossible that the rather trivial value
$X_{\infty}=0$ which is more typical of a low-temperature phase might be 
a peculiarity of the spherical model. On the other hand, if $\alpha=-2$, then
$X_{\infty}=1$ and the system evolves towards equilibrium. 
\item If the equilibrium state is in the mean field regime and if in addition
the initial correlations decay slower than in equilibrium, the 
fluctuation-dissipation ratio $X$ vanishes in the scaling limit, independently
of the value of $x=t/s$ (regimes II/V). Despite of having fixed the temperature
at its critical value, this behaviour is typical for a quench into the
low-temperature ordered phase. 
\end{enumerate}
It would be of interest to see whether a similar variety of different types
of non-equilibrium critical dynamics may be established for different spin 
systems (especially with $z\ne 2$) and in particular, 
whether the r\^ole of the equilibrium and/or the
initial state being in the mean-field regime can be confirmed. As a
preparation for this we  briefly treat in the appendix an example where 
the mean magnetization does not vanish. 

\item 
Turning to the values of the autocorrelation exponent $\lambda_C$ and the
autoresponse exponent $\lambda_R$, we find that one of the following two
scenarios is realized: either (i) the two exponents are equal
\BEQ
\lambda_C = \lambda_R
\EEQ
which occurs at criticality in the regimes III and IV characterized by
$\alpha_{\rm eff}=0$ and where also 
$X_{\infty}\ne 0$ or else (ii) they satisfy the relation
\BEQ \label{4:lClRa}
\lambda_C = \lambda_R + \alpha_{\rm eff} 
\EEQ
which is realized in the entire ordered phase and for the 
critical regimes I, II and V, where $\alpha_{\rm eff}=\alpha$ and  
$X_{\infty}=0$. As an extreme case, this also 
includes a completely ordered initial state,
where $C_{\rm ini}(\vec{r})\sim \mbox{\rm cste.}$ which corresponds to 
$\alpha=-d$. Up to eventual logarithmic corrections, this case may be included
by taking the limit $D\to 2+$. 

We point out that the results 
$\lambda_C/z=\eta/4$ and $\lambda_R/z=1+\eta/4$ obtained for the $2D$ XY model
\cite{Bert01} for a fully ordered initial state (therefore $\alpha=-d=-2$) 
in the low-temperature phase are consistent with eq.~(\ref{4:lClRa}), 
since $z=2$ in that model. Indeed, the behaviour of the fluctuation-dissipation
ratio $X_{2D\:{\rm XY}}(x)$ of the $2D$ XY model is quite analogous to the one 
found here for the spherical model in regime I at criticality. In 
figure~\ref{Abb2}b we show the fluctuation-dissipation ratio of the 
$2D$ XY model when starting form a fully ordered initial state 
\BEQ \label{4:XY2D}
X_{2D\:{\rm XY}}(x) = \left[1-\frac{(x-1)^2}{2(x+1)}\right]^{-1}
\EEQ 
which is valid in the entire low-temperature phase, 
using the spin-wave approximation \cite{Bert01}. The similarity with the
functions $X(x)$ of the spherical model in regime I with 
$\alpha<-2$ is evident. 

Therefore, we conjecture that the relationship between $\lambda_C$ and
$\lambda_R$ is in general given by eq.~(\ref{4:lClRa}) and that only in those
special cases where $\alpha_{\rm eff}=0$ applies these two exponents happen 
to be equal. That was indeed the case in all models reviewed in \cite{Godr01}.  
Tests of the conjectured scaling relation (\ref{4:lClRa}) in other models would 
be most welcome.
 
Similarly, the usual anticipation, see e.g. \cite{Godr01}, 
that at criticality the two exponents 
$a=b=2\beta/\nu z$, where $\beta$ and $\nu$ are standard equilibrium critical
exponents, can be checked through the entries of table~\ref{tab2}. It appears
again more as a property of certain initial conditions than as a generally
valid statement. From table~\ref{tab2}, $a=b$ appears to hold whenever $X(t,s)$
does not vanish identically for all $x$ in the scaling limit. In these cases
the proposed relationship to the exponents $\beta,\nu,z$ seems rather to be a
hyperscaling relation since it does not hold in regime IV. 

\item 
Having examined the asymptotic properties of the scaling functions
$f_C(x)$ and $f_R(x)$ for $x$ large, we now turn to their functional form
for finite values of $x$. Indeed, as already mentioned in the introduction,
conformal invariance predicts that the scaling function of the autoresponse
function should be given by \cite{Henk01,Henk02}
\BEQ 
f_R(x) = x^{1+a-\lambda_R/z} \left(x-1\right)^{-1-a}
\EEQ
That prediction can be tested by comparing with the exact result
(\ref{3:fRskal}) for a critical quench and with (\ref{3:Tbas1}) for a quench 
into the orderd phase. Inserting the values of the exponents $a$ and
$\lambda_R$ which can be read from table~\ref{tab2} and from 
eqs.~(\ref{3:Tbas1},\ref{3:Tbas2}), respectively, show perfect agreement,
both below criticality and at criticality for all five aging regimes. 

In addition, we may also test the full space-time dependent response function.
Since the dynamical exponent $z=2$ in our model, conformal invariance
predicts for $z=2$ \cite{Henk01,Henk94}
\BEQ \label{4:Rfull}
R_{\vec{r}}(t,s) = R(t,s) \exp\left(-\frac{\cal M}{2}\frac{r^{2}}{t-s}\right)
\EEQ
where spatial translation invariance is already taken into accout,
the autoresponse function $R(t,s)$ is given as before by eq.~(\ref{1:ConfR}) 
and $\cal M$ is a non-universal and dimensionful constant. To check this, 
we calculate the full response function from eq.~(\ref{2:Rqts}). In the scaling
limit we are interested in, where both $t,s$ as well as their difference
$t-s$ become simulataneously large, we may use the $q\to 0$ limit
(\ref{2:omega-eta}) in the integral 
\BEQ
R_{\vec{r}}(t,s) =  (2\pi)^{-d} 
\sqrt{\frac{g(T,s)}{g(T,t)}} \int\!\D\vec{q}\: 
\exp\left[-\omega(\vec{q})(t-s) - \II \vec{q}\cdot\vec{r} \right] 
\EEQ
and it is easy to see that the resulting gaussian integrals reproduce 
eq.~(\ref{4:Rfull}) exactly, with ${\cal M}=1/2$. 

This suggests that the presence of conformal invariance in aging systems
should be independent of spatially long-ranged correlations in the initial
state. Further tests of conformal invariance in different aging
systems with spatially long-range initial conditions are called for.  
\end{enumerate}

Summarising, the study of the influence of spatially long-range correlations
in the initial state on the long-time behaviour of two-time observables 
of the exactly solvable
spherical model has led us to the recognition of several new types of
non-equilibrium critical dynamics in that model. As a consequence, we could 
formulate the conjecture eq.~(\ref{4:lClRa}) on the relationship between the 
autocorrelation exponent $\lambda_C$ and the autoresponse exponent $\lambda_R$. 
Finally, the hypothesis of conformal invariance in aging ferromagnetic systems 
could be tested in a new way. 
\newpage 

\appsection{Some remarks on the $1D$ Glauber-Ising model}

Here we consider briefly some non fully disordered initial conditions in
the $1D$ kinetic Ising model with Glauber \cite{Glau63} dynamics. The 
Hamiltonian is on a chain of $N$ sites with periodic boundary conditions
\BEQ
{\cal H} = - J \sum_{i=1}^{N} \sigma_i \sigma_{i+1}
\EEQ
where $\sigma_i=\pm 1$ are the Ising spins. The dynamics may be given through
the heat bath rule, which gives the probability of finding the spin variables
$\sigma_{i}(t+1)$ in terms of those at time $t$
\BEQ
P\left(\sigma_i(t+1) = \pm 1\right) = \frac{1}{2} \left[
1\pm \tanh\left( \frac{J}{T} \left(\sigma_{i-1}(t)+\sigma_{i+1}(t)\right)
\right) \right] 
\EEQ
Here we consider the following initial conditions in terms of the spin-spin
correlator $C_{\vec{r}}(t)=\langle\sigma_{\vec{r}}(t)
\sigma_{\vec{0}}(t)\rangle$
\BEQ
C_{\vec{r}}(0) = M_0^2 + 
\left( 1 - M_0^2\right)\delta_{\vec{r},\vec{0}}
\EEQ
where $M_0=M_{\vec{r}}(0)$ is the initial averaged magnetization. This
allows a simple case study of situations where the mean magnetization
$M_{\vec{r}}(t)$ does not vanish. Here and
in the sequel spatial translation invariance is already taken into account.
We study the long-time evolution of the $1D$ model after a quench to zero
temperature at time $t=0$.  

The exact solution of the model is closely parallel to the standard lines
as presented for the special case $M_0=0$ in \cite{Godr00a,Lipp00}. 
We shall therefore merely state our
results. First, the mean magnetization 
$M_{\vec{r}}(t)=M_{\vec{r}}(0)=M_0$
does remain constant for all times $t$. 
We are interested in the {\em connected} autocorrelation 
function $\Gamma(t,s)$ and the autoresponse function $R(t,s)$ defined as
\BEA
\Gamma(t,s) &=& \langle \sigma_{\vec{r}}(t) \sigma_{\vec{r}}(s) \rangle -
\langle\sigma_{\vec{r}}(t)\rangle\langle \sigma_{\vec{r}}(s) \rangle \\
R(t,s) &=& T \left.\frac{\delta M_{\vec{r}}(t)}
{\delta H_{\vec{r}}(s)}\right|_{H_{\vec{r}}=0}
\EEA
where $H_{\vec{r}}$ is an external magnetic field at the site $\vec{r}$. 

In $1D$, we find in the scaling limit $t,s\to\infty$ 
but with $x=t/s> 1$ fixed
\BEQ
\Gamma(t,s) =  \left(1-M_0^2\right)\: \frac{2}{\pi}
\arctan\sqrt{\frac{2}{x-1}\,} \;\; , \;\;
R(t,s) = \left(1-M_0^2\right)\: \frac{1}{\pi s} 
\sqrt{\frac{1}{2(x-1)}\,}
\EEQ
and the fluctuation-dissipation ratio 
$X(t,s) = R(t,s)(\partial\Gamma(t,s)/\partial s)^{-1}$
becomes a function of $x$ only and reads
$X(t,s)=X(x)=(x+1)/2x$. Up to the prefactor 
$1-M_{\vec{r}}(t)M_{\vec{r}}(s)=1-M_0^2$, the results for
$\Gamma(t,s)$ and $R(t,s)$ (and therefore also for $X(x)$) 
are exactly the same as found in 
\cite{Godr00a,Lipp00} for the autocorrelation and autoreponse functions 
at $M_0=0$. Therefore,
unless $M_0=1$ and the system is prepared in a zero-temperature equilibrium
state, the exponent $\alpha$ introduced in the text takes the 
value $\alpha_{\rm eff}=0$ here. That agrees with the fact that the 
autocorrelation and autoresponse exponents are equal, $\lambda_C=\lambda_R=1$. 

It would be interesting to study the effects of non fully disordered initial
conditions in wider settings, e.g. in higher dimensions, conserved order
parameters and so on. 
\zeile{2}
\noindent {\large\bf Acknowledgements}\\

\noindent
MH thanks the Grup da mat\'eria condensada ao Complexo Interdisciplinar da
Universidade de Lisboa for warm hospitality, where this work was written 
up.   



}


{\small
\begin{thebibliography}{999}
\bibitem{Bray94} A.J. Bray, Adv. Phys. {\bf 43}, 357 (1994).
\bibitem{Bouc98} J.P. Bouchaud, L.F. Cugliandolo, J. Kurchan and M. M\'ezard,
in A.P. Young (ed.) {\it Spin Glasses and Random Fields},
World Scientific (Singapore 1998); ({\tt cond-mat/9702070}).
\bibitem{Cate00} M.E. Cates and M.R. Evans (eds), {\it Soft and Fragile Matter},
Proc. 53$^{\rm rd}$ Scottish University Summer Schools in Physics, 
(Bristol 2000). 
\bibitem{Godr01} C. Godr\`eche and J.M. Luck, J. Phys. Cond. Matt. {\bf 14}, 
1589 (2002).
\bibitem{Cugl94a} L.F. Cugliandolo and J. Kurchan, J. Phys. {\bf A27}, 
5749 (1994). 
\bibitem{Cugl94b} L.F. Cugliandolo, J. Kurchan, and G. Parisi, J. Physique 
{\bf I4}, 1641 (1994).
\bibitem{Zipp00} W. Zippold, R. K\"uhn and H. Horner, Eur. Phys. J. {\bf B13},
531 (2000). 
\bibitem{Fish88} D.S. Fisher and D.A. Huse, Phys. Rev. {\bf B38}, 373 (1988).
\bibitem{Huse89} D.A. Huse, Phys. Rev. {\bf B40}, 304 (1989). 
\bibitem{Godr00a} C. Godr\`eche and J.M. Luck, J. Phys. {\bf A33}, 1151 (2000).
\bibitem{Lipp00} E. Lippiello and M. Zanetti, Phys. Rev. {\bf E61}, 3369 (2000).
\bibitem{Cugl95} L.F. Cugliandolo and D.S. Dean, J. Phys. {\bf A28}, 
4213 (1995).
\bibitem{Godr00b} C. Godr\`eche and J.M. Luck, J. Phys. {\bf A33}, 9141 (2000).
\bibitem{Bert01} L. Berthier, P.C.W. Holdsworth, and M. Sellitto,
J. Phys. {\bf A34}, 1805 (2001).
\bibitem{Zote02} V.S. Zotev, G.F. Rodriguez, R. Orbach, E. Vincent and 
J. Hammann, {\tt cond-mat/0202269}. 
\bibitem{Bray91} A.J. Bray, K. Humayun and T.J. Newman, Phys. Rev. {\bf B43}, 
3699 (1991).
\bibitem{Henk99} M. Henkel, {\it Phase Transitions and Conformal Invariance},
Springer (Heidelberg 1999).
\bibitem{Henk01} M. Henkel, M. Pleimling, C. Godr\`eche and J.-M. Luck,
Phys. Rev. Lett. {\bf 87}, 265701 (2001). 
\bibitem{Henk02} M. Henkel, {\tt hep-th/0205256}. 
\bibitem{Jans89} H.K. Janssen, B. Schaub and B. Schmittmann, 
Z. Phys. {\bf B73}, 539 (1989).
\bibitem{Newm90} T.J. Newman and A.J. Bray, J. Phys. {\bf A23}, 4491 (1990).
\bibitem{Kiss93} J.G. Kissner and A.J. Bray, J. Phys. {\bf A26}, 1571 (1993). 
\bibitem{Coni94} A. Coniglio, P. Ruggiero and M. Zanetti, Phys. Rev. {\bf E50},
1046 (1994). 
\bibitem{Cala02} P. Calabrese and A. Gambassi, {\tt cond-mat/0203096}. 
\bibitem{Cann01} S.A. Cannas, D.A. Stariolo and F.A. Tamarit, Physica 
{\bf A294}, 362 (2001).
\bibitem{Corb02} F. Corberi, E. Lippiello and M. Zanetti, 
{\tt cond-mat/0202091}. 
\bibitem{Abra65} M.A. Abramowitz and I.A. Stegun, {\it Handbook of Mathematical
Functions}, Dover (New York 1965)
\bibitem{Mont65} E.W. Montroll and G.H. Weiss, J. Math. Phys. {\bf 6},
167 (1965).
\bibitem{Henk94} M. Henkel, J. Stat. Phys. {\bf 75}, 1023 (1994). 
\bibitem{Glau63} R.J. Glauber, J. Math. Phys. {\bf 4}, 294 (1963).
\end{thebibliography}
}
\end{document}